\documentclass[sigconf, screen, nonacm]{acmart}
\AtBeginDocument{%
  }

\setcopyright{acmlicensed}
\copyrightyear{2025}
\acmYear{2025}
\acmDOI{XXXXXXX.XXXXXXX}
\acmConference[Conference acronym 'XX]{Make sure to enter the correct
  conference title from your rights confirmation email}{Oct 27--31,
  2025}{Dublin}
\acmISBN{978-1-4503-XXXX-X/2018/06}

\usepackage{booktabs}
\usepackage{multirow}
\begin{document}

\title{Enhancing Non-Core Language Instruction-Following in Speech LLMs via Semi-Implicit Cross-Lingual CoT Reasoning}

\author{Hongfei Xue}
\email{hfxue@mial.nwpu.edu.cn}
\affiliation{%
  \institution{Northwestern Polytechnical University}
  \city{Xi'an}
  \country{China}
}

\author{Yufeng Tang}
\email{tangyufeng.aladdin@bytedance.com}
\affiliation{%
  \institution{ByteDance Inc.}
  \city{Beijing}
  \country{China}
}

\author{Hexin Liu}
\email{hexin.liu@ntu.edu.sg}
\affiliation{%
  \institution{Nanyang Technological University}
  \country{Singapore}
}

\author{Jun Zhang}
\email{zhangjun.jarry@bytedance.com}
\affiliation{%
  \institution{ByteDance Inc.}
  \city{Beijing}
  \country{China}
}
\author{Xuelong Geng}
\email{xlgeng@mail.nwpu.edu.cn}
\affiliation{%
  \institution{Northwestern Polytechnical University}
  \city{Xi'an}
  \country{China}
}
\author{Lei Xie}
\email{lxie@nwpu.edu.cn}
\affiliation{%
  \institution{Northwestern Polytechnical University}
  \city{Xi'an}
  \country{China}
}


\begin{abstract}
Large language models have been extended to the speech domain, leading to the development of speech large language models (SLLMs). While existing SLLMs demonstrate strong performance in speech instruction-following for core languages (e.g., English), they often struggle with non-core languages due to the scarcity of paired speech-text data and limited multilingual semantic reasoning capabilities.
To address this, we propose the semi-implicit \textbf{C}ross-lingual \textbf{S}peech \textbf{C}hain-\textbf{o}f-\textbf{T}hought (XS-CoT) framework, which integrates speech-to-text translation into the reasoning process of SLLMs. The XS-CoT generates four types of tokens: instruction and response tokens in both core and non-core languages, enabling cross-lingual transfer of reasoning capabilities. 
To mitigate inference latency in generating target non-core response tokens, we incorporate a semi-implicit CoT scheme into XS-CoT, which progressively compresses the first three types of intermediate reasoning tokens while retaining global reasoning logic during training. By leveraging the robust reasoning capabilities of the core language, XS-CoT improves responses for non-core languages by up to 45\% in GPT-4 score when compared to direct supervised fine-tuning on two representative SLLMs, Qwen2-Audio and SALMONN. Moreover, the semi-implicit XS-CoT reduces token delay by more than 50\% with a slight drop in GPT-4 scores. Importantly, XS-CoT requires only a small amount of high-quality training data for non-core languages by leveraging the reasoning capabilities of core languages. To support training, we also develop a data pipeline and open-source speech instruction-following datasets in Japanese, German, and French.




\end{abstract}




\maketitle

\section{Introduction}
Large language models (LLMs) have recently made significant progress in the field of artificial intelligence~\cite{openai2023gpt4, bai2023qwen, LLaMA, abdin2024phi, team2024gemini, guo2025deepseek}. In response to the growing demand for speech-based interactions, researchers have extended LLMs to the speech modality, resulting in the development of speech large language models (SLLMs)~\cite{23Pengi, tang2023salmonn, gong2023listentu, 24wavllm, chu2024qwen2audio, peng2024survey}. 
Unlike traditional cascade systems that treat speech recognition and natural language processing as separate tasks, SLLMs integrate these processes, enabling more natural and efficient communication with machines. 

SLLMs now demonstrate excellent capabilities in understanding and processing diverse speech signals as well as performing complex reasoning tasks~\cite{23Pengi,  gong2023listentu, tang2023salmonn, chu2024qwen2audio, 24echat}. In high-resource or core language scenarios like English (\textit{en}) and Chinese (\textit{zh}), existing SLLMs, such as Qwen2-Audio~\cite{chu2024qwen2audio}, SALMONN~\cite{tang2023salmonn}, and WavLLM~\cite{24wavllm} have achieved performance comparable to or even surpassing that of traditional automatic speech recognition (ASR) systems and text-based LLMs in speech understanding. A typical SLLM architecture comprises three core components: a speech encoder for modality-specific feature extraction, an LLM for text generation, and an adapter that bridges the gap between speech and text modalities. 
For example, SALMONN~\cite{tang2023salmonn} uses a Q-former~\cite{23Qformer} adapter to combine the Vicuna text-based LLM~\cite{chiang2023vicuna} with dual speech and audio encoders, while Qwen-Audio~\cite{chu2023qwenaudio} and Qwen2-Audio~\cite{chu2024qwen2audio} integrate the Whisper~\cite{23whisper} speech encoder with the Qwen text-based LLM~\cite{bai2023qwen} to handle various speech tasks.

Despite these advances, the development of SLLMs suffers from a significant imbalance in language resources~\cite{manakul2024enhancing}. Therefore, existing SLLMs commonly exhibit lower response quality in speech interactions involving non-core languages like Japanese compared to core languages such as English. Additionally, the LLM component within SLLMs is typically pre-trained predominantly on core languages~\cite{23CrossLingual, abdin2024phi, 24MLLMS, wendler2024llamas}. Thereby limiting the broader applicability of SLLM-based technologies. To improve the reasoning capability of LLMs in non-core languages, recent works have proposed translating non-core language inputs into a core language English with a chain-of-thought~(CoT) prompt before generating responses in the original language~\cite{23NotAllLanguage, 24DoMultilingual}. However, directly applying this strategy to SLLMs is infeasible, as SLLMs take speech signals as input rather than text, making conventional CoT prompting incompatible.

\begin{figure}[t]
  \centering
  \includegraphics[width=0.9\linewidth]{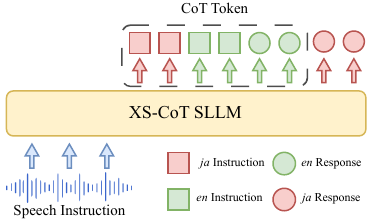}
  \caption{An overview of XS-CoT SLLM framework: speech instruction as input, text tokens as output, with red for target non-core languages (\textit{ja}) and green for core languages (\textit{en}).}
  \label{fig: overview}
  \vspace{-18pt}
\end{figure}

In this work, we focus on the task of speech instruction-following (SI), where models must understand spoken commands and execute corresponding actions.
Inspired by the CoT prompting, we propose a novel framework called \textbf{C}ross-lingual \textbf{S}peech \textbf{CoT} (XS-CoT) to improve SLLMs performance on low-resource and non-core languages. Unlike conventional CoT prompting, which is applied directly to the textual input of LLMs, the proposed XS-CoT framework embeds the speech-to-text translation process within the reasoning process of SLLMs. 
As shown in Figure~\ref{fig: overview}, the XS-CoT SLLM generates four types of text tokens sequentially: first, the instruction tokens in the target language, followed by both instruction and response tokens in the core language, and finally, the response tokens in the target language.
This enables the cross-lingual transfer of reasoning capabilities from high-resource to low-resource or non-core languages. Experimental evaluations on two representative SLLMs, SALMONN~\cite{tang2023salmonn} and Qwen2Audio~\cite{chu2024qwen2audio}, reveal that the proposed XS-CoT enhances response quality for non-core languages by approximately 45\% relative to direct supervised fine-tuning (SFT).


While the XS-CoT approach benefits non-core language processing, its multi-step nature introduces token delays and increases target language response latency. Previous works~\cite{24FromE2I, 24ThinkBeforeSpeak, yuen2024internalizing} tackled this issue through implicit CoT reasoning, which removes explicit intermediate reasoning tokens. While this approach reduces latency, it compromises the model’s ability to perform detailed semantic reasoning.
To address this trade-off, we propose a semi-implicit CoT scheme that stepwise compresses the local details of the first three types of tokens (CoT token) during training. By retaining the global CoT logic while reasoning the missing local details, our approach reduces latency by over 50\%, with only a slight drop in GPT-4 scores. 
Moreover, since the pre-trained LLM inherently possesses reasoning ability, fine-tuning for non-core languages requires only a modest amount of high-quality training data. To support this, we also introduce a data pipeline designed to generate high-quality instruction-following data and release the XS-CoT instruction-following datasets for three non-core languages: Japanese, German, and French.

The key contributions of our work are summarized as follows:
\begin{itemize}
    \item \textbf{Extending Translation-based CoT to Speech:} We propose a method that directly generates four types of tokens, including instruction and response tokens in both core and non-core languages, improving SI performance for non-core language by 45\% compared to direct SFT.
    \item \textbf{Semi-implicit XS-CoT:} Our semi-implicit CoT method stepwise compresses the first three types of tokens during training, preserving global reasoning while reducing the local details. This approach cuts the first token delay in the target language response tokens by over 50\%.
    \item \textbf{Non-core Data Generation and Open-sourcing:} We introduce a dedicated data pipeline that generates high-quality SI data, alleviating data scarcity and enabling the open-sourcing of data for Japanese, French, and German.
\end{itemize}
\vspace{-10pt}

\section{Related Work}
\subsection{Multilingual LLMs}
The increasing globalization of technology has spurred significant interest in multilingual large language models (MLLMs)~\cite{24MLLMS}. Most MLLMs leverage multilingual pre-training~\cite{21mt5, 23bloom, LLaMA} to acquire cross-lingual capabilities. However, due to the uneven distribution of pre-training data, these models tend to favor high-resource languages, resulting in unbalanced language performance~\cite{abdin2024phi, 24DataSelect, wendler2024llamas}. 

To address these limitations without incurring the high cost of pre-training from scratch, two primary approaches have been proposed to enhance multilingual performance in existing models. First, prompt-based methods utilize the model's inherent translation capabilities to convert non-core language inputs into a core language and then generate responses~\cite{23FredaShi, 23NotAllLanguage, 23CrossLingual, 23IsTranslation, 24DoMultilingual}. For example, the self-translate method~\cite{24DoMultilingual} first translates the input text into a core language during the initial inference step and subsequently generates responses in the original language, leveraging richer linguistic resources. 
The second strategy uses post-training techniques to inject language-specific knowledge into pre-trained models. This includes strategies such as continuous pre-training~\cite{24TeachingLlama, 23FinGPT} and instruction fine-tuning~\cite{23CrosslingualGeneralization, 24QuestionTranslation}. For instance, Zhu et al.~\cite{24QuestionTranslation} demonstrated that cross-language instruction fine-tuning combined with translation training can effectively achieve language alignment.

In contrast to these existing approaches, our work introduces CoT techniques into the speech modality. We extend the CoT framework to handle speech input by constructing a novel four-token output framework. Additionally, we combine ideas from both prompt-based and post-training approaches to enhance the multilingual capabilities of SLLMs through fine-tuning on a dataset enriched with XS-CoT annotations.

\subsection{Latent Reasoning in LLMs}
CoT reasoning methods have been demonstrated to improve response quality by explicitly modeling intermediate reasoning steps~\cite{22CoT, guo2025deepseek}. However, such explicit reasoning introduces significant delays for the first token, which impedes real-time applications. To mitigate this issue, recent research has explored latent reasoning approaches in LLMs.

\begin{figure*}[t]
    \centering
    \includegraphics[width=0.90\linewidth]{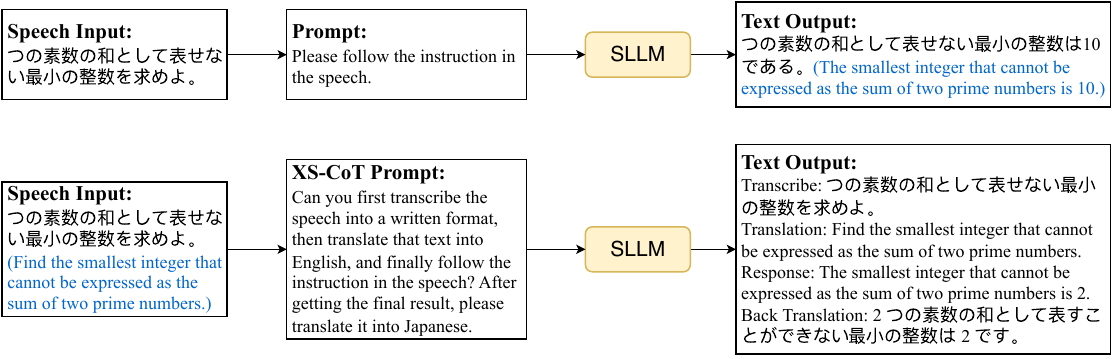}
    \caption{Example responses: direct SFT target language (\textit{ja}) output (top) vs. XS-CoT output (bottom). The words in blue are translations made for ease of understanding and are not the output of SLLM.}
    \label{fig:Example-responses}
    \vspace{-5pt}
\end{figure*}

Some prior works have attempted to augment the model with additional tokens. For instance, methods that insert learnable \texttt{<pause>} or filler tokens during training~\cite{24ThinkBeforeSpeak, 24ThinkDot} have been explored. However, Pfau et al.~\cite{24ThinkDot} argue that these methods do not enhance the expressive power of the model and may not scale to more complex reasoning tasks. 
Other approaches, such as implicit CoT reasoning~\cite{23ImplicitCoT, 24FromE2I, yuen2024internalizing}, aim to internalize the reasoning process via knowledge distillation and specialized training regimes. Deng et al.~\cite{24FromE2I} used curriculum learning to gradually remove intermediate CoT tokens while preserving reasoning performance, demonstrating effectiveness in solving math problems.
Nevertheless, applying these methods to SI is challenging. 
In the XS-CoT framework, where responses are typically lengthy, completely internalizing or removing intermediate reasoning tokens can compromise the semantic richness of the generated content. An alternative approach involves using continuous hidden layer markers in place of discrete tokens~\cite{24ContinuousLatent}, which helps to preserve the reasoning logic partially but at the cost of increased training resource consumption.

In contrast, our method retains the global logic of the core language reasoning while selectively distilling or omitting only the local details of the reasoning tokens. This approach ensures that the overall coherence of the reasoning chain is maintained during response tokens into the target language, thereby reducing the first-token delay without sacrificing response quality.

\section{Method}
Our objective is to enhance the non-core speech instruction-following capability of SLLMs by building upon existing models. We select SALMONN~\cite{tang2023salmonn} and Qwen2Audio~\cite{chu2024qwen2audio} as the baseline and further train them using our proposed Semi-Implicit XS-CoT framework.

\subsection{XS-CoT}
To address the challenge of weak semantic reasoning in non-core languages, we propose the Cross-lingual Speech CoT framework. As illustrated in Figure~\ref{fig: overview}, the output of our framework for speech instruction-following in the target language (\textit{ja}) involves the generation of four types of text tokens:

\begin{itemize}
    \item \textbf{Target Language Instruction Token.}
    The input speech instruction is first transcribed into text instruction using the SLLM fine-tuned on the target language's ASR data. This token aligns the input speech modality with the target language.

    \item \textbf{Core Language Instruction Token.}
    Given the limited support for the target non-core language, the transcribed text instruction token is translated into a core language (\textit{en}) to exploit the robust capabilities of the LLM.
    
    \item \textbf{Core Language Response Token.}
    The model performs semantic reasoning on the translated core language instruction token, generating high-quality response tokens in the core language.

    \item \textbf{Target Language Response Token.}  
    Finally, the core language response token is translated back into the target language to produce the final response.

\end{itemize}
These four types of tokens work together to produce a high-quality target response.
For example, Figure~\ref{fig:Example-responses} shows that while a direct SFT output in Japanese yields an incorrect response (\textit{``The smallest integer that cannot be expressed as the sum of two prime numbers is 10"}), the XS-CoT output correctly states (\textit{``The smallest integer that cannot be expressed as the sum of two prime numbers is 2"}).

\subsection{Semi-Implicit XS-CoT}
\begin{figure*}[t]
  \centering
  \includegraphics[width=1.0\linewidth]{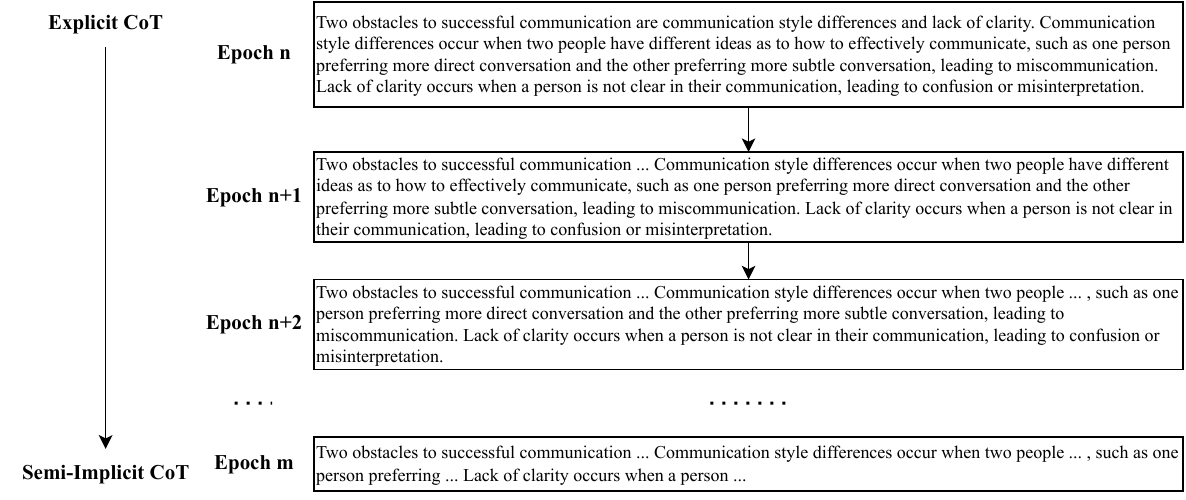}
  \caption{Stepwise internalization for semi-implicit XS-CoT reasoning in core language response token. The process progressively reduces the number of tokens during training, thereby decreasing inference latency.}
  \label{fig:Semi-implicit-CoT}
  \vspace{-10pt}
\end{figure*}

Explicit CoT reasoning enhances response quality by generating intermediate reasoning tokens but incurs substantial token delay. Prior work on implicit CoT~\cite{23ImplicitCoT, 24FromE2I, yuen2024internalizing} addresses this by internalizing these tokens; however, fully internalizing the reasoning chain can degrade performance, especially for long, complex responses.
To balance reasoning quality and latency, we propose a \textbf{Semi-Implicit CoT} approach for XS-CoT. Our method preserves key intermediate tokens while compressing less critical details through a two-level segmentation. Specifically, we perform both sentence segmentation and word segmentation on the full reasoning chain.
The full reasoning chain is divided into sentences, allowing us to retain the global semantic structure. Within each sentence, only the first \(k\) groups of words are retained, with an ellipsis (\texttt{...}) appended to indicate omitted details. 
Word segmentation can keep the semantics of the first groups of tokens relatively complete and reduce independent tokens.
This strategy reduces verbosity while maintaining the overall reasoning chain. Figure~\ref{fig:Semi-implicit-CoT} illustrates the stepwise internalization process for the core language response token.

During training, we gradually increase the degree of internalization. Let \(x\) be the number of sentences in the segmented reasoning chain. At training epoch \(n\),  we internalize the minimum number of first few sentences between \(x\) and \(n\) (with each sentence compressed as described). Once training reaches a pre-defined epoch \(m\), all \(x\) sentences are fully internalized. Formally, we define the number of internalized sentences as follows:
\begin{align}
\text{Internalized Sentences} = 
\begin{cases}
    \min(x, n)  & \text{if } n < m, \\
    x  & \text{if } n \geq m.
\end{cases}
\end{align}

To ensure training stability, we employ an optimizer reset strategy after each epoch to prevent abrupt gradient changes caused by loss mutations~\cite{24FromE2I}. Additionally, with a probability \(p\), an extra sentence is internalized at each epoch, which helps the model adapt smoothly to the evolving internalization process.

We experiment with two internalization strategies.
The first one is to internalize all three CoT tokens, including the target language instruction token and both instruction and response tokens in the core language. Since instruction tokens are brief, they are fully internalized in one epoch, while the response token in the core language is progressively internalized.
To further refine the model's understanding of instructions, we also experiment with internalizing only the longest and most complex response token in the core language while retaining the complete token for others.

\begin{figure}[h]
  \centering
  \includegraphics[width=1.0\linewidth]{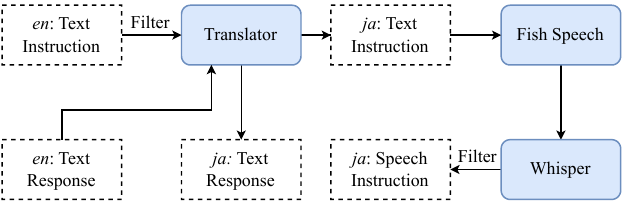}
  \caption{Overview of the data pipeline for generating four types of tokens. `\textit{en}' stands for English, `\textit{ja}' stands for the target language Japanese.}
  \label{fig:data pipline}
  \vspace{-10pt}
\end{figure}

\subsection{Training Stages}
\label{sec: training}
We adopt a three-stage training strategy to improve the SLLM’s target non-core language speech instruction-following capability.

\begin{itemize}
    \item \textbf{Stage 1: Modal Alignment.}  
    For target languages that the SLLM does not originally support, we fine-tune the model using a large corpus of target language ASR data, while retaining a portion of the original data to preserve performance on previous tasks.
    
    \item \textbf{Stage 2: XS-CoT Fine-tuning.}  
    Once the model reliably transcribes target language speech, we further fine-tune it using our XS-CoT dataset, which implements the complete four types of tokens. A smaller portion of original task data is included to maintain balance.

    \item \textbf{Stage 3: Semi-implicit CoT Training.}  
    With high-quality responses already achievable by XS-CoT, we apply semi-implicit training to compress and internalize the reasoning tokens, thereby reducing token latency without significantly sacrificing response quality.
    
\end{itemize}
After these stages, the SLLM is capable of generating high-quality target language responses with reduced latency.

\subsection{Data Pipline}
To support our training, we propose a data pipeline for generating target language XS-CoT data, as depicted in Figure~\ref{fig:data pipline}.
We start with Alpaca text instruction data\footnote{\url{https://github.com/tatsu-lab/stanford\_alpaca}}~\cite{alpaca}, and filter out instruction-response pairs that are either uncommon or noisy for speech applications.
The filtered text data is translated into the target language using the internal translation API.
The translated instruction-response pairs are then converted into speech instructions using an open-source TTS model fish-speech v1.5\footnote{\url{https://github.com/fishaudio/fish-speech}}~\cite{fishspeech}. This model supports multiple speaker variations per language to generate diverse and natural-sounding data.
Additionally, an ASR model\footnote{\url{https://huggingface.co/openai/whisper-large-v3}}~\cite{23whisper} is used to evaluate the synthesized speech, and only samples with a word error rate (WER) below 5\% are retained.    
Using this pipeline, we have generated and open-sourced high-quality speech instruction-following data (Multilingual-Alpaca-Speech) for \textit{ja}, \textit{fr}, and \textit{de}.

\begin{table}[]
\caption{Training data used in the experiment. For non-core languages, Multilingual Alpaca Speech includes dataset for direct SFT and dataset for XS-CoT.}
\label{tab: Datasets}
\begin{tabular}{@{}cccc@{}}
\toprule
Task                 & Data Source       & Hours (h) & Samples (k) \\ \midrule
\multirow{4}{*}{ASR} & LibriSpeech (\textit{en})~\cite{15librispeech}  & 960     & 280     \\
                     & Reazonspeech (\textit{ja})~\cite{fujimoto2016reazonspeech} & 1000    & -         \\
                     & Multilingual LibriSpeech (\textit{fr})~\cite{20mllibrispeech}          & 1000    & -         \\
                     & Multilingual LibriSpeech (\textit{de})~\cite{20mllibrispeech}          & 1000    & -         \\
AST                  & CoVoST2 (\textit{en2zh})~\cite{21covost2}   & 430     & 290     \\
SQA                  & LibriSpeech~\cite{tang2023salmonn}       & 960     & 280     \\
PR                   & LibriSpeech~\cite{tang2023salmonn}       & 960     & 280     \\
GR                   & LibriSpeech~\cite{tang2023salmonn}       & 100     & 28       \\
\multirow{4}{*}{SI}  & Multilingual Alpaca Speech (\textit{en})       & -       & 60       \\
                     & Multilingual Alpaca Speech (\textit{ja})       & -       & 30       \\
                     & Multilingual Alpaca Speech (\textit{fr})       & -       & 10       \\
                     & Multilingual Alpaca Speech (\textit{de})       & -       & 10       \\ \bottomrule
\end{tabular}
\vspace{-10pt}
\end{table}

\begin{table*}[]
\caption{Main Results in terms of GPT-4 scores $\uparrow$ and average generated token length $\downarrow$. `Tin' represents text instruction token, `Tout' represents text response token. `S' represents speech. () represents the average number of all generated tokens from LLM.}
\begin{tabular}{@{}lllcccc@{}}
\toprule
{}   & {}   & {}   & \multicolumn{2}{c}{{SALMONN}}  & \multicolumn{2}{c}{{Qwen2Audio}}      \\
\multirow{-2}{*}{{Exp}} & \multirow{-2}{*}{{Model}} & \multirow{-2}{*}{{\{\} represents CoT}}   & {OpenHermes} & \multicolumn{1}{l}{{ALPACA}} & {OpenHermes} & \multicolumn{1}{l}{{ALPACA}} \\ \midrule
{e1}  & {LLM}  & {\textit{en} Tin → \textit{en} Tout}    & {69.6 (216)}            & {73.8 (130)}                            & {72.2 (243)}            & {73.2 (144)}                            \\
{e2}  & {SLLM} & {\textit{en} Sin → \textit{en} Tout}       & {58.0 (72)}             & {68.4 (48)}                             & {59.8 (133)}            & {64.0 (60)}                           \\
{e3}  & {LLM}  & {\textit{ja} Tin → \textit{ja} Tout}                 & {49.2 (468)}            & {45.0 (331)}                            & {47.1 (360)}            & {44.2 (265)}                            \\
{e4}  & {LLM + CoT} & \textit{ja} Tin → \{\textit{en Tin → en Tout}\} → \textit{ja Tout}            & {51.6 (441)}            & {51.6 (289)}                            & {50.0 (418)}            & {48.0 (344)}                            \\
{e5}  & {SLLM} & {\textit{ja} Sin → \textit{ja} Tout}        & {26.3 (181)}            & {30.4 (140)}                            & {39.4 (160)}            & {40.6 (124)}                            \\
{e6}  & {SLLM + XS-CoT} & {\begin{tabular}[c]{@{}l@{}}\textit{ja} Sin → \{\textit{ja Tin → en Tin → en Tout}\}  → \textit{ja} Tout\end{tabular}}                   & {49.4 (262)}   & {51.2  (179)}                   & {46.5 (289)}   & {51.0 (193)}                    \\
{e7}  & {\begin{tabular}[c]{@{}l@{}}SLLM + Semi-implict\end{tabular}} & {\begin{tabular}[c]{@{}l@{}}\textit{ja} Sin → \{\textit{ja Tin → en Tin → en Tout}\}  → \textit{ja} Tout\end{tabular}}  & {40.2 (190)}    & {45.8 (143)}                    & {44.0 (261)}    & {43.8 (160)}                    \\ 
\bottomrule
\end{tabular}
\label{tab: main results}
\vspace{-10pt}
\end{table*}

\section{Experiment Setup}
\subsection{Datasets}
Table~\ref{tab: Datasets} provides an overview of the datasets used in our experiments. 
We use English (\textit{en}) as the core language and Japanese (\textit{ja}), German (\textit{de}) and French (\textit{fr}) as non-core languages.
We adopt a three-stage training strategy (described in Section~\ref{sec: training}) to enhance the speech instruction-following capability of SLLMs in non-core languages.
In the stage 1 modal alignment, we use target language ASR data and SI data to achieve modality alignment. For example, we use the Reazonspeech~\cite{fujimoto2016reazonspeech} training set and direct SFT data of the Multilingual Alpaca Speech for Japanese. To preserve the model's performance on original tasks, we also include data such as LibriSpeech~\cite{15librispeech} for ASR, CoVoST2~\cite{21covost2} for AST, and LibriSpeech-based datasets~\cite{tang2023salmonn} for speech question answering (SQA), phone recognition (PR), and gender recognition (GR).
In the stage 2 XS-CoT fine-tuning, 
we sample a subset of the stage 1 data and augment it with our generated XS-CoT format data of the Multilingual Alpaca Speech training set, which is designed explicitly for our XS-CoT process.
In the final stage of semi-implicit CoT training, we use training data similar to stage 2 to refine the model parameters further, reducing token latency while optimizing performance on target language tasks.

\subsection{Implementation Details}
We validate our method by extending the target language capabilities on two widely used SLLMs, SALMONN~\cite{tang2023salmonn} and Qwen2Audio~\cite{chu2024qwen2audio}. Each model undergoes the three-stage training process with the following configurations.

\textbf{SALMONN:}  
We follow the original model settings~\cite{tang2023salmonn} by freezing both speech encoders and training only the Q-former and the LLM using the LoRA adapter~\cite{22lora}.  
Given the relatively weak SI performance (even in English), we first fine-tune on the Multilingual Alpaca Speech (\textit{en}) data to obtain the SALMONN-instruction-7b version.
Thereafter, we then apply the three-stage training process.
For stage 1, only the Q-former is fine-tuned on 8*A100 GPUs, with a peak learning rate of 2e-5, batch size 16, and 30,000 steps (with a 10\% warmup). 
In stages 2 and 3, only the LLM is fine-tuned using LoRA, with a peak learning rate of 1e-5, batch size 16, and 10,000 steps.
For stage 3, the hyperparameter $m$ is set to 7 (comprising one epoch for target instruction token, one for core instruction token, and 5 for core response token), $p$ is 0.1, and $k$ is 7.

\textbf{Qwen2Audio:}  
We use the ms-swift~\cite{zhao2024swiftascalablelightweightinfrastructure} framework to implement the three-stage training. 
In stage 1, only the encoder and adapter are updated for modality alignment using 8*A100 GPUs, a peak learning rate of 2e-5, batch size 16, and 30,000 steps. 
In stages 2 and 3, we fine-tune only the LLM with LoRA (peak learning rate of 1e-5, batch size 16, and 10,000 steps). For Stage 3, $m$ is set to 5 (core language response token only), with $p$ at 0.1 and $k$ at 3.

\subsection{Evaluation Metrics}
We evaluate our models using AudioBench\footnote{\url{https://github.com/AudioLLMs/AudioBench/tree/main}}~\cite{wang2024audiobench}, which comprises eight tasks across 26 datasets focusing on speech understanding, interpretation, and audio scene understanding. For our experiments, we focus on tasks related to speech understanding: ASR, AST, SQA, and SI.
To construct the target non-core language SI test set, we translate the SI test set from AudioBench using a reliable translation system and synthesize the corresponding speech with fish-speech v1.5. This process yields two SI test sets for the target language: OpenHermes and ALPACA.
For the SI and SQA tasks, we use the GPT-4 score metric to evaluate performance. Following AudioBench, the model's predictions and the ground truth are provided to GPT-4, which assigns a score from 0 to 5. These scores are then scaled to a 0-100 range to comparison with previous work.

\begin{table}[t]
\caption{Latent reasoning results measured by GPT-4 score $\uparrow$ and generated CoT token length $\downarrow$. <> represents the number of the CoT token. Note that CoT tokens do not contain responses in the target language.}
\begin{tabular}{@{}lccc@{}}
\toprule
\multicolumn{1}{l}{{Exp}} & {OpenHermes} & \multicolumn{1}{c}{{ALPACA}} & \multicolumn{1}{c}{{Avg}} \\ \midrule
{Whisper+\textit{ja}-llama-13b}    & {57.1 {<}0{>} }                  & {50.6 {<}0{>}} & {53.9 {<}0{>}}                                  \\
{Direct SFT}              & {26.3 {<}0{>}}          & {30.4 {<}0{>}}       & {28.4 {<}0{>}}                   \\
{XS-CoT}            & {49.4 {<}128{>}}        & {51.2 {<}85{>}}       & {50.3 {<}107{>}}                  \\
{\texttt{<pause>} tokens~\cite{24ThinkBeforeSpeak}}               & {29.6 {<}16{>}}         & {30.2 {<}13{>}} & {29.9 {<}15{>}}                        \\
{Implicit CoT~\cite{24FromE2I, yuen2024internalizing}}            & {28.0 {<}0{>}}          & {28.6 {<}0{>}}        & {28.3 {<}0{>}}                  \\
{Semi-Implicit XS-CoT}       & {40.2 {<}56{>}}         & {45.8 {<}49{>}}        & {43.0 {<}53{>}}                 \\ \bottomrule
\end{tabular}
\label{tab: latent reasoning}
\vspace{-10pt}
\end{table}

\section{Experiment Results}

\begin{table*}[]
\caption{Performance of XS-CoT on Original SLLM Tasks. Although slight regressions are observed in some tasks, overall functionality remains unaffected.}
\begin{tabular}{@{}lcccccc@{}}
\toprule
& \multicolumn{2}{c}{ASR (WER $\downarrow$)}   & AST (BLEU $\uparrow$) & \multicolumn{2}{c}{SI (GPT-4 Score $\uparrow$)}    & SQA (GPT-4 Score $\uparrow$)                                   \\ \midrule
{}                       & {LibriSpeech (\textit{en})} & \multicolumn{1}{l}{{CommonVoice (\textit{ja})}} & {CoVoST2 (\textit{en2zh})} & {OpenHermes (\textit{en})} & {ALPACA (\textit{en})} & {slue\_p2\_sqa5 (\textit{en})} \\ \midrule
{SALMONN} & {5.36}          & {-}                                    & {34.7}    & {58.0}            & {68.4}        & {81.8}           \\
{XS-CoT}                 & {6.81}          & {7.48}                                 & {28.5}    & {56.4}            & {66.2}        & {81.1}           \\ \bottomrule
\end{tabular}
\label{tab: original tasks}
\vspace{-5pt}
\end{table*}

\begin{table*}[]
\label{tab:multilingual}
\caption{Multilingual capabilities results measured by GPT-4 score $\uparrow$. <> represents the number of the CoT token $\downarrow$, including the firt three types of intermediate tokens. The hyperparameter \(k\) is set to 3.}
\begin{tabular}{@{}lllllllll@{}}
\toprule
                                 & \multicolumn{4}{c}{OpenHermes}                   & \multicolumn{4}{c}{ALPACA}                     \\ \midrule
                                 & \textit{de}             & \textit{fr}             & \textit{ja}      & Avg       & \textit{de}             & \textit{fr}            & \textit{ja}    & Avg        \\ \midrule
Text-based LLM                     & 55.4 {<}0{>}          & 55.2 {<}0{>}          & 49.2 {<}0{>}   & 53.3 {<}0{>}        & 52.2 {<}0{>}          & 53.4 {<}0{>}         & 45.0 {<}0{>}     & 50.2 {<}0{>}         \\
Direct SFT      & 31.2 {<}0{>}   & 36.8 {<}0{>}   & 16.1 {<}0{>} & 28.0 {<}0{>}  & 32.6 {<}0{>}   & 35.6 {<}0{>}  & 24.4 {<}0{>} & 30.9 {<}0{>} \\
XS-CoT     & 51.2 {<}135{>} & 48.8 {<}133{>} & 42.0 {<}143{>}  & 47.3 {<}137{>} & 43.8 {<}103{>} & 52.4 {<}94{>} & 53.4 {<}99{>}  & 49.9 {<}99{>} \\
Semi-Implicit XS-CoT & 38.6 {<}29{>}  & 40.4 {<}31{>}  & 32.2 {<}31{>} & 37.1 {<}30{>}  & 34.6 {<}27{>}  & 40.0 {<}28{>} & 36.8 {<}30{>}  & 37.1 {<}28{>} \\ \bottomrule
\end{tabular}
\label{tab: multilingual results}
\vspace{-10pt}
\end{table*}

\subsection{Extending Single Language}

\textbf{Main Results}
We select \textit{ja} as the target non-core language and conduct training on both SALMONN~\cite{tang2023salmonn} and Qwen2Audio~\cite{chu2024qwen2audio}. Models are compared with two types of instruction inputs: text-based instructions (LLM) and speech-based instructions (SLLM). The experimental conditions are defined as follows:
\begin{itemize}
    \item \textbf{e1:} Original text-based LLM models of SALMONN (Vicuna-7B~\cite{chiang2023vicuna}) and Qwen2Audio (Qwen-7B~\cite{bai2023qwen}) evaluated with \textit{en} text instructions.
    \item \textbf{e2:} The SLLM models of newly trained SALMONN-instruction-7B and original Qwen2Audio-7B evaluated with \textit{en} speech instructions.
    \item \textbf{e3:} Models from e1, but evaluated with \textit{ja} text instructions to assess response quality in Japanese.
    \item \textbf{e4:} Same as e3, but incorporating a CoT reasoning prompt.
    \item \textbf{e5:} Baseline model trained with direct SFT using Multilingual Alpaca Speech \textit{ja}.
    \item \textbf{e6:} Model trained with our XS-CoT method (Stage 2).
    \item \textbf{e7:} Model trained with our Semi-Implicit XS-CoT method (Stage 3).
    \item \textbf{Whisper + \textit{ja}-based LLM:} As a topline reference (see Table~\ref{tab: latent reasoning}), we also include a cascade system consisting of the Whisper ASR model and a text-based LLM~\footnote{\url{https://huggingface.co/llm-jp/llm-jp-3-13b-instruct}} trained specifically for Japanese.
\end{itemize}
Our analysis reveals several key observations based on the results in Table~\ref{tab: main results}.
\textbf{(1) Modality Conversion Loss:} 
In e2, GPT-4 scores are consistently lower than in e1 across four test sets. This performance drop is primarily due to information loss during the conversion from text to speech instructions. Similarly, e5 underperforms compared to e3, indicating that direct SFT with Multilingual Alpaca Speech (\textit{ja}) results in lower-quality responses. Additionally, speech-based instructions lead to shorter average generated token length, which may be related to speech SFT data, further compounding performance degradation. This highlights the challenges posed not only by modality alignment but also by shorter output length in SLLMs.
\textbf{(2) Benefits of XS-CoT:} The improvements in GPT-4 scores from e3 to e4 and from e5 to e6 demonstrate the positive impact of incorporating CoT reasoning. Notably, in the speech input scenario (e6), XS-CoT nearly matches or even exceeds the scores of the text-based LLM e4, benefiting its fine-tuning on XS-CoT data. The SALMONN and Qwen2Audio models achieve an average relative improvement of approximately 45\% on GPT-4 scores compared to the baseline (e5). 
In Table~\ref{tab: latent reasoning}, the XS-CoT results for SALMONN show an average GPT-4 score of 50.3, which is even comparable to the topline performance of the cascaded system of ``Whisper+\textit{ja}-llama''.
This indicates that the XS-CoT training method significantly enhances performance. 
\textbf{(3) Effectiveness of Semi-Implict CoT:}
Comparing e6 with e7, the Semi-Implicit XS-CoT reduces the total number of average generated tokens by compressing the reasoning tokens. The performance of e7 falls between e5 and e6, suggesting that our method is effective across both models. In the following section, we will further analyze Semi-Implicit CoT in more detail.

\noindent\textbf{Semi-Implict CoT}
To analyze the performance and advantages of the Semi-Implicit CoT method, we conduct a detailed experiment, with results summarized in Table~\ref{tab: latent reasoning}. We compare our method with two representative approaches in implicit reasoning. (1) Additional Token~\cite{24ThinkBeforeSpeak}: This approach introduces special learnable tokens into the reasoning process. In our framework, each sentence is replaced by a \texttt{<pause>} token during the step-by-step internalization training. (2) Implicit CoT~\cite{24FromE2I, yuen2024internalizing}: This method adopts a more radical strategy by completely eliminating all tokens from the reasoning process, thereby relying solely on internalized reasoning without any explicit token guidance.

The experimental results in Table~\ref{tab: latent reasoning} reveal that although both the additional \texttt{<pause>} token and the implicit CoT significantly reduce or eliminate the reasoning chains, they yield only limited improvements in performance compared to the baseline model trained directly via SFT. We attribute this to the inherent complexity of generating a long and logically coherent response in the target language.
In contrast, our proposed Semi-Implicit XS-CoT method retains the essential global reasoning structure while effectively reducing the number of inference tokens. Specifically, compared to the direct SFT model, the Semi-Implicit approach achieves a 51.4\% relative improvement in GPT-4 score. Although it registers a 14.5\% lower GPT-4 score relative to the full XS-CoT method, it reduces the CoT token count required for reasoning by 50\%. This trade-off demonstrates that our method successfully balances efficiency and performance, offering a practical solution for enhancing both the speed and quality of complex reasoning tasks.


\noindent\textbf{Original Tasks}
While the XS-CoT method improves target language capabilities, it is crucial to evaluate its impact on original tasks. Table~\ref{tab: original tasks} compares the SALMONN-instruction-7b model with the XS-CoT trained model across several tasks.
For \textit{en}-based tasks such as ASR, SI, and SQA, performance remains consistent between the original and XS-CoT trained models. This consistency suggests that the XS-CoT training process, which also includes a reasoning component in the core language (\textit{en}), effectively preserves the original model's competencies.
In contrast, the model shows a slight drop in performance on the \textit{en2zh} of the AST task. This drop may be attributed to the fact that there are fewer Chinese outputs of the LLM during training, which places more emphasis on the target language.
Notably, the XS-CoT trained model demonstrates enhanced new capabilities, especially in target language ASR and SI tasks. These improvements underscore the benefits of the XS-CoT training method in bolstering the model's multilingual processing abilities without severely compromising its original functions.

\begin{figure}[t]
  \centering
  \includegraphics[width=1.0\linewidth]{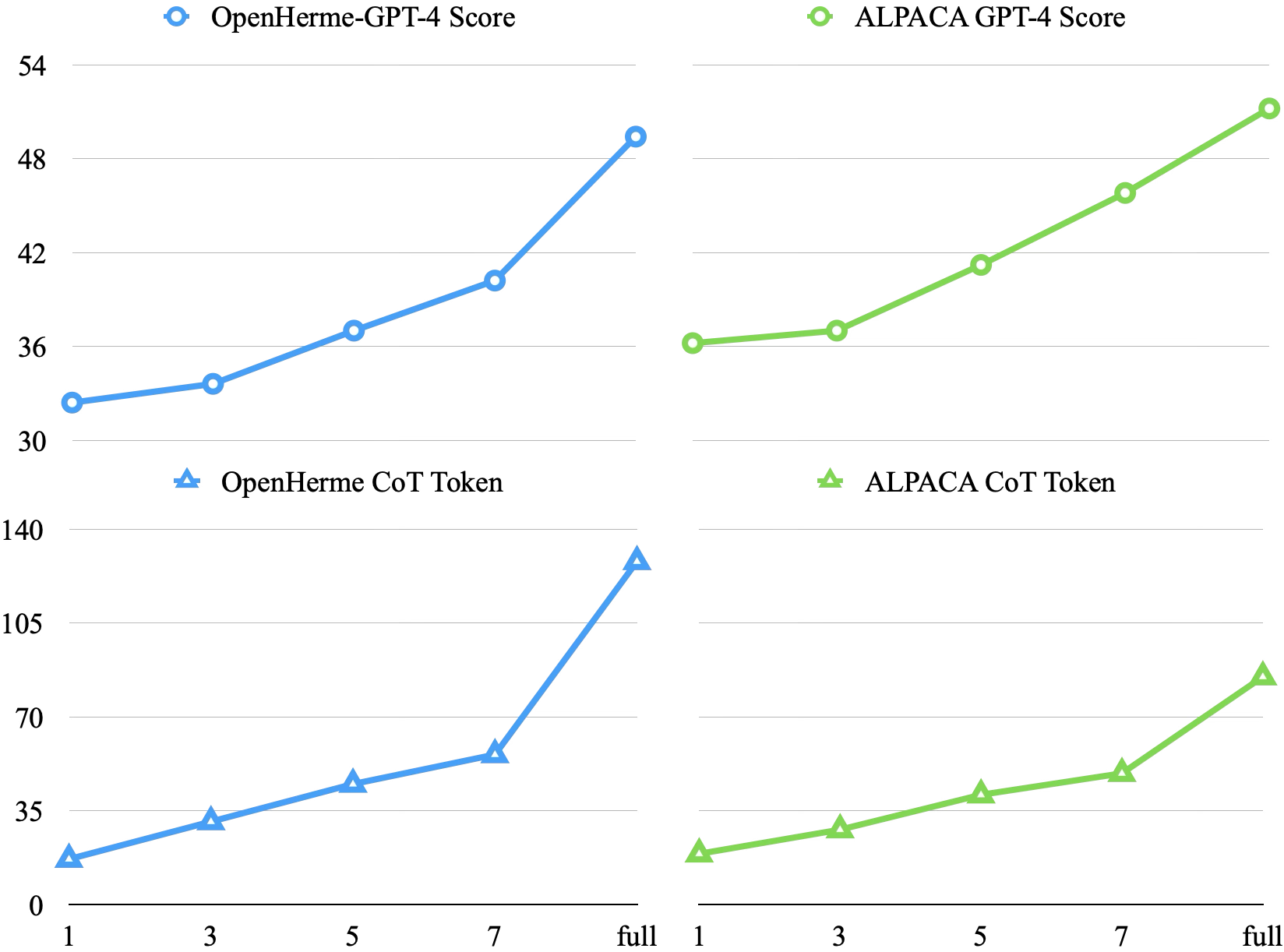}
  \caption{Impact of the hyperparameter \(k\) on the Semi-Implicit CoT method. The horizontal axis represents different \(k\) values (1, 3, 5, 7, and full reasoning chain), with the top sub-graphs displaying GPT-4 scores and the bottom sub-graphs showing the average generated CoT token lengths.}
  \label{fig:k-index}
  \vspace{-15pt}
\end{figure}

\noindent\textbf{Retain the First \(k\) Groups of Words}
We examine the influence of the hyperparameter \(k\) on Semi-Implicit CoT performance. Figure~\ref{fig:k-index} presents the results, with the horizontal axis representing different \(k\) values (1, 3, 5, 7, and “full,” representing the complete reasoning chain without token reduction). The top sub-graphs show GPT-4 scores, and the bottom sub-graphs show average CoT token lengths.
As expected, increasing \(k\) leads to better GPT-4 scores, suggesting that more reasoning tokens contribute to more coherent and complete responses. However, the number of CoT tokens also increases, reflecting the additional computational cost of a more detailed reasoning process. These trends confirm our hypothesis that while higher \(k\) values improve performance, they come with an increased inference cost. Therefore, this improvement must be balanced against the increased inference cost due to the longer token sequences.

\vspace{-10pt}
\subsection{Extending Multiple Language}

\begin{figure*}[t]
  \centering
  \includegraphics[width=0.86\linewidth]{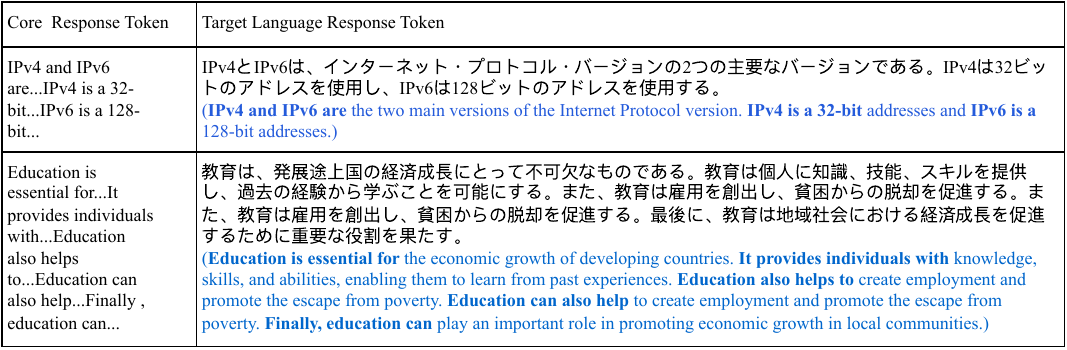}
  \caption{Comparison of Semi-Implicit core language response token with target language response token. The words in blue are translations made for ease of understanding and are not the output of the SLLM.}
  \label{fig: Samples}
  \vspace{-15pt}
\end{figure*}

In this experiment, we evaluate the scalability of our model in a multilingual setting by testing it on German (\textit{de}), French (\textit{fr}), and Japanese (\textit{ja}). Our goal is to assess the universality and effectiveness of our model across various languages. The experiment is conducted using the SALMONN-instruction-7b model with continuous training, and the results are summarized in Table~\ref{tab: multilingual results}. The training process is divided into three stages.
Stage 1: The training data is augmented with ASR and SI data for \textit{de}, \textit{fr} and \textit{ja}, alongside the basic dataset. This stage establishes multilingual capabilities.
Stage 2: To ensure a balanced comparison, each language is supplemented with 10k Multilingual Alpaca Speech samples. This enhancement focuses on improving the model's reasoning and instruction-following abilities for each language.
Stage 3: We set the key hyperparameter \( k \) to 3, which controls the extent of information distillation and internalization during semi-implicit training. This stage refines the model's performance, focusing on both efficiency and quality.

The results indicate that after XS-CoT training, all three languages show significant performance improvements in speech instruction-following tasks. Compared to the model fine-tuned via SFT, the GPT-4 scores improved by an average of 19.3 points on the OpenHermes test set and 19 points on ALPACA. Notably, the performance in speech input scenarios matches that of direct text input to text-based LLM, demonstrating robust performance.
Furthermore, after applying the step-by-step internalization and distillation of the Semi-Implicit method, the model's GPT-4 score falls between the direct SFT baseline and the full XS-CoT results. By setting the internalization hyperparameter \( k \) to 3, we achieve a substantial reduction in the delay of the first token, from approximately 137 tokens to just 30 tokens, greatly enhancing real-time response capability.

\subsection{Ablation Study}
To assess the contributions of different reasoning steps in our proposed method, we conduct ablation studies by progressively removing key components of the XS-CoT framework. Additionally, we analyze the impact of the Semi-Implicit internalization process.

\begin{table}[t]
\caption{XS-CoT Ablation Results. <> represents the number of the CoT token.}
\begin{tabular}{@{}lccc@{}}
\toprule
    & OpenHermes      & ALPACA        & Avg       \\ \midrule
XS-CoT & 49.4 {<}128{>} & 51.2 {<}85{>}  & 50.3 {<}107{>}  \\
- \textit{ja} Instruction & 42.4 {<}81{>}  & 46.4 {<}56{>}  & 44.4 {<}67{>} \\
\ \ - \textit{en} Instruction & 37.6 {<}80{>}  & 41.6 {<}50{>}  & 39.6 {<}65{>} \\ 
\ \ \ \ - \textit{en} Response & 26.3 {<}0{>}  & 30.4 {<}50{>} & 28.4 {<}0{>}  \\ \bottomrule
\end{tabular}
\label{tab: XS-CoT Ablation}
\vspace{-10pt}
\end{table}

\begin{table}[t]
\caption{Semi-Implicit CoT Ablation (with \(k = 3\)). <> represents the number of the CoT token.}
\begin{tabular}{@{}lccc@{}}
\toprule
& OpenHermes                     & ALPACA        & Avg                 \\ \midrule
Semi-Implicit XS-CoT     & {33.6 {<}31{>}} & {37.0 {<}28{>}} & {35.3 {<}30{>}} \\
- \textit{ja} Instruction     & {36.6 {<}65{>}} & {39.4 {<}50{>}} & {38.0 {<}58{>}} \\
\ \ - \textit{en} Instruction & {38.8 {<}77{>}} & {45.4 {<}57{>}}  & {42.1 {<}67{>}} \\
\ \ \ \ - \textit{en} Response & {49.4 {<}128{>}} & {51.2 {<}85{>}}  & {50.3 {<}107{>}} \\ \bottomrule
\end{tabular}
\label{tab: Semi-implicit-ablation}
\vspace{-10pt}
\end{table}

\textbf{XS-CoT}
Table~\ref{tab: XS-CoT Ablation} presents the ablation study of the XS-CoT process, where we sequentially remove the reasoning steps for the first three types of tokens.
The results demonstrate that as more reasoning steps are omitted, the number of CoT tokens required in intermediate reasoning decreases. However, this reduction comes at the cost of a significant drop in response quality. This confirms that each type of tokens in XS-CoT plays an essential role in maintaining high-quality outputs.
Notably, in the first one type and two types of tokens are removed, performance declines substantially, yet the number of inference tokens does not decrease proportionally. We attribute this to the absence of explicit instruction text, particularly in first two types of tokens. Direct removal of \textit{en} instructions from \textit{ja} speech inputs to \textit{en} responses creates a large semantic gap, making it difficult for the LLM to reason the response accurately. This results in unstable fluctuations in the number of response tokens.


\textbf{Semi-Implicit CoT}
Table~\ref{tab: Semi-implicit-ablation} shows the results of the Semi-Implicit CoT ablation, where the internalization level of the first three types of tokens is progressively reduced.
The results indicate that as internalization decreases, the number of tokens required in the reasoning process gradually increases while the GPT-4 evaluation scores also improve. This suggests a trade-off that deeper internalization reduces the computational cost and token usage but at the expense of response quality.

\subsection{Qualitative Analysis of Semi-Implicit CoT}
To assess whether the Semi-Implicit CoT method retains global reasoning and infers missing local details, we present two representative examples in Figure~\ref{fig: Samples}.  
In the first example, consider the intermediate CoT reasoning text: "\textit{IPv4 and IPv6 are... IPv4 is a 32-bit...}". 
When generating the final target language response tokens, the model retains the core semantics of the original sentence through the initial word groups of each clause. 
Additionally, for the ellipsis-represented content ("..."), the model successfully infers contextually plausible details by leveraging the global reasoning flow.
In another scenario, the model also demonstrates robust continuative reasoning for core language response tokens. However, when generating content for core response tokens like "\textit{Education also helps to...}", which spans two sentences, the model's outputs tend to be overly similar and lack diversity. We attribute this limitation to the model's relatively weaker processing abilities in the target non-core language, resulting in less varied output.

\section{Conclusion}
In this paper, we introduced a novel semi-implicit cross-lingual speech chain-of-thought (XS-CoT) framework to improve instruction-following capabilities in speech LLMs for non-core languages. By decomposing the process into four sequential types of tokens, the proposed method leverages the robust reasoning capabilities of core languages to enhance the response quality of non-core languages. Additionally, our proposed semi-implicit CoT method uses sentence segmentation and stepwise compression to mitigate the token delay introduced by explicit reasoning chains. We also develop a dedicated data pipeline that generates high-quality speech instruction data, as well as open-sourced datasets in Japanese, German, and French. These advancements contribute to more effective and efficient speech LLMs for non-core languages. 



\clearpage
\bibliographystyle{ACM-Reference-Format}
\bibliography{refs}


\begin{thebibliography}{48}


\ifx \showCODEN    \undefined \def \showCODEN     #1{\unskip}     \fi
\ifx \showISBNx    \undefined \def \showISBNx     #1{\unskip}     \fi
\ifx \showISBNxiii \undefined \def \showISBNxiii  #1{\unskip}     \fi
\ifx \showISSN     \undefined \def \showISSN      #1{\unskip}     \fi
\ifx \showLCCN     \undefined \def \showLCCN      #1{\unskip}     \fi
\ifx \shownote     \undefined \def \shownote      #1{#1}          \fi
\ifx \showarticletitle \undefined \def \showarticletitle #1{#1}   \fi
\ifx \showURL      \undefined \def \showURL       {\relax}        \fi
\providecommand\bibfield[2]{#2}
\providecommand\bibinfo[2]{#2}
\providecommand\natexlab[1]{#1}
\providecommand\showeprint[2][]{arXiv:#2}

\bibitem[Abdin et~al\mbox{.}(2024)]%
        {abdin2024phi}
\bibfield{author}{\bibinfo{person}{Marah Abdin}, \bibinfo{person}{Sam~Ade Jacobs}, \bibinfo{person}{Ammar~Ahmad Awan}, \bibinfo{person}{Jyoti Aneja}, \bibinfo{person}{Ahmed Awadallah}, \bibinfo{person}{Hany Awadalla}, \bibinfo{person}{Nguyen Bach}, \bibinfo{person}{Amit Bahree}, \bibinfo{person}{Arash Bakhtiari}, \bibinfo{person}{Harkirat Behl}, {et~al\mbox{.}}} \bibinfo{year}{2024}\natexlab{}.
\newblock \showarticletitle{Phi-3 technical report: A highly capable language model locally on your phone}.
\newblock \bibinfo{journal}{\emph{arXiv preprint arXiv:2404.14219}} (\bibinfo{year}{2024}).
\newblock


\bibitem[Albalak et~al\mbox{.}(2024)]%
        {24DataSelect}
\bibfield{author}{\bibinfo{person}{Alon Albalak}, \bibinfo{person}{Yanai Elazar}, \bibinfo{person}{Sang~Michael Xie}, \bibinfo{person}{Shayne Longpre}, \bibinfo{person}{Nathan Lambert}, \bibinfo{person}{Xinyi Wang}, \bibinfo{person}{Niklas Muennighoff}, \bibinfo{person}{Bairu Hou}, \bibinfo{person}{Liangming Pan}, \bibinfo{person}{Haewon Jeong}, \bibinfo{person}{Colin Raffel}, \bibinfo{person}{Shiyu Chang}, \bibinfo{person}{Tatsunori Hashimoto}, {and} \bibinfo{person}{William~Yang Wang}.} \bibinfo{year}{2024}\natexlab{}.
\newblock \showarticletitle{A Survey on Data Selection for Language Models}.
\newblock \bibinfo{journal}{\emph{Trans. Mach. Learn. Res.}}  \bibinfo{volume}{2024} (\bibinfo{year}{2024}).
\newblock


\bibitem[Bai et~al\mbox{.}(2023)]%
        {bai2023qwen}
\bibfield{author}{\bibinfo{person}{Jinze Bai}, \bibinfo{person}{Shuai Bai}, \bibinfo{person}{Yunfei Chu}, \bibinfo{person}{Zeyu Cui}, \bibinfo{person}{Kai Dang}, \bibinfo{person}{Xiaodong Deng}, \bibinfo{person}{Yang Fan}, \bibinfo{person}{Wenbin Ge}, \bibinfo{person}{Yu Han}, \bibinfo{person}{Fei Huang}, {et~al\mbox{.}}} \bibinfo{year}{2023}\natexlab{}.
\newblock \showarticletitle{Qwen technical report}.
\newblock \bibinfo{journal}{\emph{arXiv preprint arXiv:2309.16609}} (\bibinfo{year}{2023}).
\newblock


\bibitem[Chiang et~al\mbox{.}(2023)]%
        {chiang2023vicuna}
\bibfield{author}{\bibinfo{person}{Wei-Lin Chiang}, \bibinfo{person}{Zhuohan Li}, \bibinfo{person}{Ziqing Lin}, \bibinfo{person}{Ying Sheng}, \bibinfo{person}{Zhanghao Wu}, \bibinfo{person}{Hao Zhang}, \bibinfo{person}{Lianmin Zheng}, \bibinfo{person}{Siyuan Zhuang}, \bibinfo{person}{Yonghao Zhuang}, \bibinfo{person}{Joseph~E Gonzalez}, {et~al\mbox{.}}} \bibinfo{year}{2023}\natexlab{}.
\newblock \showarticletitle{Vicuna: An open-source chatbot impressing gpt-4 with 90\%* chatgpt quality}.
\newblock \bibinfo{journal}{\emph{See https://vicuna. lmsys. org (accessed 14 April 2023)}} \bibinfo{volume}{2}, \bibinfo{number}{3} (\bibinfo{year}{2023}), \bibinfo{pages}{6}.
\newblock


\bibitem[Chu et~al\mbox{.}(2024)]%
        {chu2024qwen2audio}
\bibfield{author}{\bibinfo{person}{Yunfei Chu}, \bibinfo{person}{Jin Xu}, \bibinfo{person}{Qian Yang}, \bibinfo{person}{Haojie Wei}, \bibinfo{person}{Xipin Wei}, \bibinfo{person}{Zhifang Guo}, \bibinfo{person}{Yichong Leng}, \bibinfo{person}{Yuanjun Lv}, \bibinfo{person}{Jinzheng He}, \bibinfo{person}{Junyang Lin}, \bibinfo{person}{Chang Zhou}, {and} \bibinfo{person}{Jingren Zhou}.} \bibinfo{year}{2024}\natexlab{}.
\newblock \showarticletitle{Qwen2-Audio Technical Report}.
\newblock \bibinfo{journal}{\emph{arXiv preprint arXiv:2407.10759}} (\bibinfo{year}{2024}).
\newblock


\bibitem[Chu et~al\mbox{.}(2023)]%
        {chu2023qwenaudio}
\bibfield{author}{\bibinfo{person}{Yunfei Chu}, \bibinfo{person}{Jin Xu}, \bibinfo{person}{Xiaohuan Zhou}, \bibinfo{person}{Qian Yang}, \bibinfo{person}{Shiliang Zhang}, \bibinfo{person}{Zhijie Yan}, \bibinfo{person}{Chang Zhou}, {and} \bibinfo{person}{Jingren Zhou}.} \bibinfo{year}{2023}\natexlab{}.
\newblock \showarticletitle{Qwen-Audio: Advancing Universal Audio Understanding via Unified Large-Scale Audio-Language Models}.
\newblock \bibinfo{journal}{\emph{arXiv preprint arXiv:2311.07919}} (\bibinfo{year}{2023}).
\newblock


\bibitem[Deng et~al\mbox{.}(2024)]%
        {24FromE2I}
\bibfield{author}{\bibinfo{person}{Yuntian Deng}, \bibinfo{person}{Yejin Choi}, {and} \bibinfo{person}{Stuart~M. Shieber}.} \bibinfo{year}{2024}\natexlab{}.
\newblock \showarticletitle{From Explicit CoT to Implicit CoT: Learning to Internalize CoT Step by Step}.
\newblock \bibinfo{journal}{\emph{CoRR}}  \bibinfo{volume}{abs/2405.14838} (\bibinfo{year}{2024}).
\newblock


\bibitem[Deng et~al\mbox{.}(2023)]%
        {23ImplicitCoT}
\bibfield{author}{\bibinfo{person}{Yuntian Deng}, \bibinfo{person}{Kiran Prasad}, \bibinfo{person}{Roland Fernandez}, \bibinfo{person}{Paul Smolensky}, \bibinfo{person}{Vishrav Chaudhary}, {and} \bibinfo{person}{Stuart~M. Shieber}.} \bibinfo{year}{2023}\natexlab{}.
\newblock \showarticletitle{Implicit Chain of Thought Reasoning via Knowledge Distillation}.
\newblock \bibinfo{journal}{\emph{CoRR}}  \bibinfo{volume}{abs/2311.01460} (\bibinfo{year}{2023}).
\newblock


\bibitem[Deshmukh et~al\mbox{.}(2023)]%
        {23Pengi}
\bibfield{author}{\bibinfo{person}{Soham Deshmukh}, \bibinfo{person}{Benjamin Elizalde}, \bibinfo{person}{Rita Singh}, {and} \bibinfo{person}{Huaming Wang}.} \bibinfo{year}{2023}\natexlab{}.
\newblock \showarticletitle{Pengi: An Audio Language Model for Audio Tasks}. In \bibinfo{booktitle}{\emph{Advances in Neural Information Processing Systems 36: Annual Conference on Neural Information Processing Systems 2023, NeurIPS 2023, New Orleans, LA, USA, December 10 - 16, 2023}}.
\newblock


\bibitem[Etxaniz et~al\mbox{.}(2024)]%
        {24DoMultilingual}
\bibfield{author}{\bibinfo{person}{Julen Etxaniz}, \bibinfo{person}{Gorka Azkune}, \bibinfo{person}{Aitor Soroa}, \bibinfo{person}{Oier~Lopez de Lacalle}, {and} \bibinfo{person}{Mikel Artetxe}.} \bibinfo{year}{2024}\natexlab{}.
\newblock \showarticletitle{Do Multilingual Language Models Think Better in English?}. In \bibinfo{booktitle}{\emph{Proceedings of the 2024 Conference of the North American Chapter of the Association for Computational Linguistics: Human Language Technologies: Short Papers, {NAACL} 2024, Mexico City, Mexico, June 16-21, 2024}}. \bibinfo{publisher}{Association for Computational Linguistics}, \bibinfo{pages}{550--564}.
\newblock


\bibitem[Gong et~al\mbox{.}(2024)]%
        {gong2023listentu}
\bibfield{author}{\bibinfo{person}{Yuan Gong}, \bibinfo{person}{Hongyin Luo}, \bibinfo{person}{Alexander~H. Liu}, \bibinfo{person}{Leonid Karlinsky}, {and} \bibinfo{person}{James~R. Glass}.} \bibinfo{year}{2024}\natexlab{}.
\newblock \showarticletitle{Listen, Think, and Understand}. In \bibinfo{booktitle}{\emph{{ICLR}}}. \bibinfo{publisher}{OpenReview.net}.
\newblock


\bibitem[Goyal et~al\mbox{.}({[n.\,d.]})]%
        {24ThinkBeforeSpeak}
\bibfield{author}{\bibinfo{person}{Sachin Goyal}, \bibinfo{person}{Ziwei Ji}, \bibinfo{person}{Ankit~Singh Rawat}, \bibinfo{person}{Aditya~Krishna Menon}, \bibinfo{person}{Sanjiv Kumar}, {and} \bibinfo{person}{Vaishnavh Nagarajan}.} \bibinfo{year}{[n.\,d.]}\natexlab{}.
\newblock \showarticletitle{Think before you speak: Training Language Models With Pause Tokens}. In \bibinfo{booktitle}{\emph{The Twelfth International Conference on Learning Representations, {ICLR} 2024, Vienna, Austria, May 7-11, 2024}}. \bibinfo{publisher}{OpenReview.net}.
\newblock


\bibitem[Guo et~al\mbox{.}(2025)]%
        {guo2025deepseek}
\bibfield{author}{\bibinfo{person}{Daya Guo}, \bibinfo{person}{Dejian Yang}, \bibinfo{person}{Haowei Zhang}, \bibinfo{person}{Junxiao Song}, \bibinfo{person}{Ruoyu Zhang}, \bibinfo{person}{Runxin Xu}, \bibinfo{person}{Qihao Zhu}, \bibinfo{person}{Shirong Ma}, \bibinfo{person}{Peiyi Wang}, \bibinfo{person}{Xiao Bi}, {et~al\mbox{.}}} \bibinfo{year}{2025}\natexlab{}.
\newblock \showarticletitle{Deepseek-r1: Incentivizing reasoning capability in llms via reinforcement learning}.
\newblock \bibinfo{journal}{\emph{arXiv preprint arXiv:2501.12948}} (\bibinfo{year}{2025}).
\newblock


\bibitem[Hao et~al\mbox{.}(2024)]%
        {24ContinuousLatent}
\bibfield{author}{\bibinfo{person}{Shibo Hao}, \bibinfo{person}{Sainbayar Sukhbaatar}, \bibinfo{person}{DiJia Su}, \bibinfo{person}{Xian Li}, \bibinfo{person}{Zhiting Hu}, \bibinfo{person}{Jason Weston}, {and} \bibinfo{person}{Yuandong Tian}.} \bibinfo{year}{2024}\natexlab{}.
\newblock \showarticletitle{Training Large Language Models to Reason in a Continuous Latent Space}.
\newblock \bibinfo{journal}{\emph{CoRR}}  \bibinfo{volume}{abs/2412.06769} (\bibinfo{year}{2024}).
\newblock


\bibitem[Hu et~al\mbox{.}(2022)]%
        {22lora}
\bibfield{author}{\bibinfo{person}{Edward~J. Hu}, \bibinfo{person}{Yelong Shen}, \bibinfo{person}{Phillip Wallis}, \bibinfo{person}{Zeyuan Allen{-}Zhu}, \bibinfo{person}{Yuanzhi Li}, \bibinfo{person}{Shean Wang}, \bibinfo{person}{Lu Wang}, {and} \bibinfo{person}{Weizhu Chen}.} \bibinfo{year}{2022}\natexlab{}.
\newblock \showarticletitle{LoRA: Low-Rank Adaptation of Large Language Models}. In \bibinfo{booktitle}{\emph{{ICLR}}}. \bibinfo{publisher}{OpenReview.net}.
\newblock


\bibitem[Hu et~al\mbox{.}(2024)]%
        {24wavllm}
\bibfield{author}{\bibinfo{person}{Shujie Hu}, \bibinfo{person}{Long Zhou}, \bibinfo{person}{Shujie Liu}, \bibinfo{person}{Sanyuan Chen}, \bibinfo{person}{Lingwei Meng}, \bibinfo{person}{Hongkun Hao}, \bibinfo{person}{Jing Pan}, \bibinfo{person}{Xunying Liu}, \bibinfo{person}{Jinyu Li}, \bibinfo{person}{Sunit Sivasankaran}, \bibinfo{person}{Linquan Liu}, {and} \bibinfo{person}{Furu Wei}.} \bibinfo{year}{2024}\natexlab{}.
\newblock \showarticletitle{WavLLM: Towards Robust and Adaptive Speech Large Language Model}. In \bibinfo{booktitle}{\emph{Findings of the Association for Computational Linguistics: {EMNLP} 2024, Miami, Florida, USA, November 12-16, 2024}}. \bibinfo{publisher}{Association for Computational Linguistics}, \bibinfo{pages}{4552--4572}.
\newblock


\bibitem[Huang et~al\mbox{.}(2023)]%
        {23NotAllLanguage}
\bibfield{author}{\bibinfo{person}{Haoyang Huang}, \bibinfo{person}{Tianyi Tang}, \bibinfo{person}{Dongdong Zhang}, \bibinfo{person}{Xin Zhao}, \bibinfo{person}{Ting Song}, \bibinfo{person}{Yan Xia}, {and} \bibinfo{person}{Furu Wei}.} \bibinfo{year}{2023}\natexlab{}.
\newblock \showarticletitle{Not All Languages Are Created Equal in LLMs: Improving Multilingual Capability by Cross-Lingual-Thought Prompting}. In \bibinfo{booktitle}{\emph{Findings of the Association for Computational Linguistics: {EMNLP} 2023, Singapore, December 6-10, 2023}}. \bibinfo{publisher}{Association for Computational Linguistics}, \bibinfo{pages}{12365--12394}.
\newblock


\bibitem[Kuulmets et~al\mbox{.}(2024)]%
        {24TeachingLlama}
\bibfield{author}{\bibinfo{person}{Hele{-}Andra Kuulmets}, \bibinfo{person}{Taido Purason}, \bibinfo{person}{Agnes Luhtaru}, {and} \bibinfo{person}{Mark Fishel}.} \bibinfo{year}{2024}\natexlab{}.
\newblock \showarticletitle{Teaching Llama a New Language Through Cross-Lingual Knowledge Transfer}. In \bibinfo{booktitle}{\emph{Findings of the Association for Computational Linguistics: {NAACL} 2024, Mexico City, Mexico, June 16-21, 2024}}. \bibinfo{publisher}{Association for Computational Linguistics}, \bibinfo{pages}{3309--3325}.
\newblock


\bibitem[Li et~al\mbox{.}(2023)]%
        {23Qformer}
\bibfield{author}{\bibinfo{person}{Junnan Li}, \bibinfo{person}{Dongxu Li}, \bibinfo{person}{Silvio Savarese}, {and} \bibinfo{person}{Steven C.~H. Hoi}.} \bibinfo{year}{2023}\natexlab{}.
\newblock \showarticletitle{{BLIP-2:} Bootstrapping Language-Image Pre-training with Frozen Image Encoders and Large Language Models}. In \bibinfo{booktitle}{\emph{International Conference on Machine Learning, {ICML} 2023, 23-29 July 2023, Honolulu, Hawaii, {USA}}} \emph{(\bibinfo{series}{Proceedings of Machine Learning Research}, Vol.~\bibinfo{volume}{202})}. \bibinfo{publisher}{{PMLR}}, \bibinfo{pages}{19730--19742}.
\newblock


\bibitem[Liao et~al\mbox{.}(2024)]%
        {fishspeech}
\bibfield{author}{\bibinfo{person}{Shijia Liao}, \bibinfo{person}{Yuxuan Wang}, \bibinfo{person}{Tianyu Li}, \bibinfo{person}{Yifan Cheng}, \bibinfo{person}{Ruoyi Zhang}, \bibinfo{person}{Rongzhi Zhou}, {and} \bibinfo{person}{Yijin Xing}.} \bibinfo{year}{2024}\natexlab{}.
\newblock \bibinfo{title}{Fish-Speech: Leveraging Large Language Models for Advanced Multilingual Text-to-Speech Synthesis}.
\newblock
\showeprint[arxiv]{2411.01156}~[cs.SD]


\bibitem[Liu et~al\mbox{.}(2024)]%
        {23IsTranslation}
\bibfield{author}{\bibinfo{person}{Chaoqun Liu}, \bibinfo{person}{Wenxuan Zhang}, \bibinfo{person}{Yiran Zhao}, \bibinfo{person}{Anh~Tuan Luu}, {and} \bibinfo{person}{Lidong Bing}.} \bibinfo{year}{2024}\natexlab{}.
\newblock \showarticletitle{Is Translation All You Need? {A} Study on Solving Multilingual Tasks with Large Language Models}.
\newblock \bibinfo{journal}{\emph{CoRR}}  \bibinfo{volume}{abs/2403.10258} (\bibinfo{year}{2024}).
\newblock


\bibitem[Luukkonen et~al\mbox{.}(2023)]%
        {23FinGPT}
\bibfield{author}{\bibinfo{person}{Risto Luukkonen}, \bibinfo{person}{Ville Komulainen}, \bibinfo{person}{Jouni Luoma}, \bibinfo{person}{Anni Eskelinen}, \bibinfo{person}{Jenna Kanerva}, \bibinfo{person}{Hanna{-}Mari Kupari}, \bibinfo{person}{Filip Ginter}, \bibinfo{person}{Veronika Laippala}, \bibinfo{person}{Niklas Muennighoff}, \bibinfo{person}{Aleksandra Piktus}, \bibinfo{person}{Thomas Wang}, \bibinfo{person}{Nouamane Tazi}, \bibinfo{person}{Teven~Le Scao}, \bibinfo{person}{Thomas Wolf}, \bibinfo{person}{Osma Suominen}, \bibinfo{person}{Samuli Sairanen}, \bibinfo{person}{Mikko Merioksa}, \bibinfo{person}{Jyrki Heinonen}, \bibinfo{person}{Aija Vahtola}, \bibinfo{person}{Samuel Antao}, {and} \bibinfo{person}{Sampo Pyysalo}.} \bibinfo{year}{2023}\natexlab{}.
\newblock \showarticletitle{FinGPT: Large Generative Models for a Small Language}. In \bibinfo{booktitle}{\emph{Proceedings of the 2023 Conference on Empirical Methods in Natural Language Processing, {EMNLP} 2023, Singapore, December 6-10, 2023}}. \bibinfo{publisher}{Association for Computational Linguistics}, \bibinfo{pages}{2710--2726}.
\newblock


\bibitem[Manakul et~al\mbox{.}(2024)]%
        {manakul2024enhancing}
\bibfield{author}{\bibinfo{person}{Potsawee Manakul}, \bibinfo{person}{Guangzhi Sun}, \bibinfo{person}{Warit Sirichotedumrong}, \bibinfo{person}{Kasima Tharnpipitchai}, {and} \bibinfo{person}{Kunat Pipatanakul}.} \bibinfo{year}{2024}\natexlab{}.
\newblock \showarticletitle{Enhancing low-resource language and instruction following capabilities of audio language models}.
\newblock \bibinfo{journal}{\emph{arXiv preprint arXiv:2409.10999}} (\bibinfo{year}{2024}).
\newblock


\bibitem[Muennighoff et~al\mbox{.}(2023)]%
        {23CrosslingualGeneralization}
\bibfield{author}{\bibinfo{person}{Niklas Muennighoff}, \bibinfo{person}{Thomas Wang}, \bibinfo{person}{Lintang Sutawika}, \bibinfo{person}{Adam Roberts}, \bibinfo{person}{Stella Biderman}, \bibinfo{person}{Teven~Le Scao}, \bibinfo{person}{M.~Saiful Bari}, \bibinfo{person}{Sheng Shen}, \bibinfo{person}{Zheng~Xin Yong}, \bibinfo{person}{Hailey Schoelkopf}, \bibinfo{person}{Xiangru Tang}, \bibinfo{person}{Dragomir Radev}, \bibinfo{person}{Alham~Fikri Aji}, \bibinfo{person}{Khalid Almubarak}, \bibinfo{person}{Samuel Albanie}, \bibinfo{person}{Zaid Alyafeai}, \bibinfo{person}{Albert Webson}, \bibinfo{person}{Edward Raff}, {and} \bibinfo{person}{Colin Raffel}.} \bibinfo{year}{2023}\natexlab{}.
\newblock \showarticletitle{Crosslingual Generalization through Multitask Finetuning}. In \bibinfo{booktitle}{\emph{Proceedings of the 61st Annual Meeting of the Association for Computational Linguistics (Volume 1: Long Papers), {ACL} 2023, Toronto, Canada, July 9-14, 2023}}. \bibinfo{publisher}{Association for Computational Linguistics}, \bibinfo{pages}{15991--16111}.
\newblock


\bibitem[OpenAI(2023)]%
        {openai2023gpt4}
\bibfield{author}{\bibinfo{person}{OpenAI}.} \bibinfo{year}{2023}\natexlab{}.
\newblock \showarticletitle{GPT-4 Technical Report}.
\newblock \bibinfo{journal}{\emph{arXiv preprint arXiv:2308.11276}} (\bibinfo{year}{2023}).
\newblock


\bibitem[Panayotov et~al\mbox{.}(2015)]%
        {15librispeech}
\bibfield{author}{\bibinfo{person}{Vassil Panayotov}, \bibinfo{person}{Guoguo Chen}, \bibinfo{person}{Daniel Povey}, {and} \bibinfo{person}{Sanjeev Khudanpur}.} \bibinfo{year}{2015}\natexlab{}.
\newblock \showarticletitle{Librispeech: An {ASR} corpus based on public domain audio books}. In \bibinfo{booktitle}{\emph{2015 {IEEE} International Conference on Acoustics, Speech and Signal Processing, {ICASSP} 2015, South Brisbane, Queensland, Australia, April 19-24, 2015}}. \bibinfo{publisher}{{IEEE}}, \bibinfo{pages}{5206--5210}.
\newblock


\bibitem[Peng et~al\mbox{.}(2024)]%
        {peng2024survey}
\bibfield{author}{\bibinfo{person}{Jing Peng}, \bibinfo{person}{Yucheng Wang}, \bibinfo{person}{Yu Xi}, \bibinfo{person}{Xu Li}, \bibinfo{person}{Xizhuo Zhang}, {and} \bibinfo{person}{Kai Yu}.} \bibinfo{year}{2024}\natexlab{}.
\newblock \showarticletitle{A survey on speech large language models}.
\newblock \bibinfo{journal}{\emph{arXiv preprint arXiv:2410.18908}} (\bibinfo{year}{2024}).
\newblock


\bibitem[Pfau et~al\mbox{.}(2024)]%
        {24ThinkDot}
\bibfield{author}{\bibinfo{person}{Jacob Pfau}, \bibinfo{person}{William Merrill}, {and} \bibinfo{person}{Samuel~R. Bowman}.} \bibinfo{year}{2024}\natexlab{}.
\newblock \showarticletitle{Let's Think Dot by Dot: Hidden Computation in Transformer Language Models}.
\newblock \bibinfo{journal}{\emph{CoRR}}  \bibinfo{volume}{abs/2404.15758} (\bibinfo{year}{2024}).
\newblock


\bibitem[Pratap et~al\mbox{.}(2020)]%
        {20mllibrispeech}
\bibfield{author}{\bibinfo{person}{Vineel Pratap}, \bibinfo{person}{Qiantong Xu}, \bibinfo{person}{Anuroop Sriram}, \bibinfo{person}{Gabriel Synnaeve}, {and} \bibinfo{person}{Ronan Collobert}.} \bibinfo{year}{2020}\natexlab{}.
\newblock \showarticletitle{{MLS:} {A} Large-Scale Multilingual Dataset for Speech Research}. In \bibinfo{booktitle}{\emph{Interspeech}}. \bibinfo{publisher}{{ISCA}}, \bibinfo{pages}{2757--2761}.
\newblock


\bibitem[Qin et~al\mbox{.}(2023)]%
        {23CrossLingual}
\bibfield{author}{\bibinfo{person}{Libo Qin}, \bibinfo{person}{Qiguang Chen}, \bibinfo{person}{Fuxuan Wei}, \bibinfo{person}{Shijue Huang}, {and} \bibinfo{person}{Wanxiang Che}.} \bibinfo{year}{2023}\natexlab{}.
\newblock \showarticletitle{Cross-lingual Prompting: Improving Zero-shot Chain-of-Thought Reasoning across Languages}. In \bibinfo{booktitle}{\emph{Proceedings of the 2023 Conference on Empirical Methods in Natural Language Processing, {EMNLP} 2023, Singapore, December 6-10, 2023}}. \bibinfo{publisher}{Association for Computational Linguistics}, \bibinfo{pages}{2695--2709}.
\newblock


\bibitem[Qin et~al\mbox{.}(2024)]%
        {24MLLMS}
\bibfield{author}{\bibinfo{person}{Libo Qin}, \bibinfo{person}{Qiguang Chen}, \bibinfo{person}{Yuhang Zhou}, \bibinfo{person}{Zhi Chen}, \bibinfo{person}{Yinghui Li}, \bibinfo{person}{Lizi Liao}, \bibinfo{person}{Min Li}, \bibinfo{person}{Wanxiang Che}, {and} \bibinfo{person}{Philip~S. Yu}.} \bibinfo{year}{2024}\natexlab{}.
\newblock \showarticletitle{Multilingual Large Language Model: {A} Survey of Resources, Taxonomy and Frontiers}.
\newblock \bibinfo{journal}{\emph{CoRR}}  \bibinfo{volume}{abs/2404.04925} (\bibinfo{year}{2024}).
\newblock


\bibitem[Radford et~al\mbox{.}(2023)]%
        {23whisper}
\bibfield{author}{\bibinfo{person}{Alec Radford}, \bibinfo{person}{Jong~Wook Kim}, \bibinfo{person}{Tao Xu}, \bibinfo{person}{Greg Brockman}, \bibinfo{person}{Christine McLeavey}, {and} \bibinfo{person}{Ilya Sutskever}.} \bibinfo{year}{2023}\natexlab{}.
\newblock \showarticletitle{Robust Speech Recognition via Large-Scale Weak Supervision}. In \bibinfo{booktitle}{\emph{ICML}}, Vol.~\bibinfo{volume}{202}. \bibinfo{pages}{28492--28518}.
\newblock


\bibitem[Scao et~al\mbox{.}(2022)]%
        {23bloom}
\bibfield{author}{\bibinfo{person}{Teven~Le Scao}, \bibinfo{person}{Angela Fan}, \bibinfo{person}{Christopher Akiki}, \bibinfo{person}{Ellie Pavlick}, \bibinfo{person}{Suzana Ilic}, \bibinfo{person}{Daniel Hesslow}, \bibinfo{person}{Roman Castagn{\'{e}}}, \bibinfo{person}{Alexandra~Sasha Luccioni}, \bibinfo{person}{Fran{\c{c}}ois Yvon}, \bibinfo{person}{Matthias Gall{\'{e}}}, \bibinfo{person}{Jonathan Tow}, \bibinfo{person}{Alexander~M. Rush}, \bibinfo{person}{Stella Biderman}, \bibinfo{person}{Albert Webson}, \bibinfo{person}{Pawan~Sasanka Ammanamanchi}, \bibinfo{person}{Thomas Wang}, \bibinfo{person}{Beno{\^{\i}}t Sagot}, \bibinfo{person}{Niklas Muennighoff}, \bibinfo{person}{Albert~Villanova del Moral}, \bibinfo{person}{Olatunji Ruwase}, \bibinfo{person}{Rachel Bawden}, \bibinfo{person}{Stas Bekman}, \bibinfo{person}{Angelina McMillan{-}Major}, \bibinfo{person}{Iz Beltagy}, \bibinfo{person}{Huu Nguyen}, \bibinfo{person}{Lucile Saulnier}, \bibinfo{person}{Samson Tan}, \bibinfo{person}{Pedro~Ortiz Suarez},
  \bibinfo{person}{Victor Sanh}, \bibinfo{person}{Hugo Lauren{\c{c}}on}, \bibinfo{person}{Yacine Jernite}, \bibinfo{person}{Julien Launay}, \bibinfo{person}{Margaret Mitchell}, \bibinfo{person}{Colin Raffel}, \bibinfo{person}{Aaron Gokaslan}, \bibinfo{person}{Adi Simhi}, \bibinfo{person}{Aitor Soroa}, \bibinfo{person}{Alham~Fikri Aji}, \bibinfo{person}{Amit Alfassy}, \bibinfo{person}{Anna Rogers}, \bibinfo{person}{Ariel~Kreisberg Nitzav}, \bibinfo{person}{Canwen Xu}, \bibinfo{person}{Chenghao Mou}, \bibinfo{person}{Chris Emezue}, \bibinfo{person}{Christopher Klamm}, \bibinfo{person}{Colin Leong}, \bibinfo{person}{Daniel van Strien}, \bibinfo{person}{David~Ifeoluwa Adelani}, {and} \bibinfo{person}{et al.}} \bibinfo{year}{2022}\natexlab{}.
\newblock \showarticletitle{{BLOOM:} {A} 176B-Parameter Open-Access Multilingual Language Model}.
\newblock \bibinfo{journal}{\emph{CoRR}}  \bibinfo{volume}{abs/2211.05100} (\bibinfo{year}{2022}).
\newblock


\bibitem[Shi et~al\mbox{.}(2023)]%
        {23FredaShi}
\bibfield{author}{\bibinfo{person}{Freda Shi}, \bibinfo{person}{Mirac Suzgun}, \bibinfo{person}{Markus Freitag}, \bibinfo{person}{Xuezhi Wang}, \bibinfo{person}{Suraj Srivats}, \bibinfo{person}{Soroush Vosoughi}, \bibinfo{person}{Hyung~Won Chung}, \bibinfo{person}{Yi Tay}, \bibinfo{person}{Sebastian Ruder}, \bibinfo{person}{Denny Zhou}, \bibinfo{person}{Dipanjan Das}, {and} \bibinfo{person}{Jason Wei}.} \bibinfo{year}{2023}\natexlab{}.
\newblock \showarticletitle{Language models are multilingual chain-of-thought reasoners}. In \bibinfo{booktitle}{\emph{The Eleventh International Conference on Learning Representations, {ICLR} 2023, Kigali, Rwanda, May 1-5, 2023}}. \bibinfo{publisher}{OpenReview.net}.
\newblock


\bibitem[Tang et~al\mbox{.}(2024)]%
        {tang2023salmonn}
\bibfield{author}{\bibinfo{person}{Changli Tang}, \bibinfo{person}{Wenyi Yu}, \bibinfo{person}{Guangzhi Sun}, \bibinfo{person}{Xianzhao Chen}, \bibinfo{person}{Tian Tan}, \bibinfo{person}{Wei Li}, \bibinfo{person}{Lu Lu}, \bibinfo{person}{Zejun Ma}, {and} \bibinfo{person}{Chao Zhang}.} \bibinfo{year}{2024}\natexlab{}.
\newblock \showarticletitle{{SALMONN:} Towards Generic Hearing Abilities for Large Language Models}. In \bibinfo{booktitle}{\emph{{ICLR}}}. \bibinfo{publisher}{OpenReview.net}.
\newblock


\bibitem[Taori et~al\mbox{.}(2023)]%
        {alpaca}
\bibfield{author}{\bibinfo{person}{Rohan Taori}, \bibinfo{person}{Ishaan Gulrajani}, \bibinfo{person}{Tianyi Zhang}, \bibinfo{person}{Yann Dubois}, \bibinfo{person}{Xuechen Li}, \bibinfo{person}{Carlos Guestrin}, \bibinfo{person}{Percy Liang}, {and} \bibinfo{person}{Tatsunori~B. Hashimoto}.} \bibinfo{year}{2023}\natexlab{}.
\newblock \bibinfo{title}{Stanford Alpaca: An Instruction-following LLaMA model}.
\newblock \bibinfo{howpublished}{\url{https://github.com/tatsu-lab/stanford_alpaca}}.
\newblock


\bibitem[Team et~al\mbox{.}(2024)]%
        {team2024gemini}
\bibfield{author}{\bibinfo{person}{Gemini Team}, \bibinfo{person}{Petko Georgiev}, \bibinfo{person}{Ving~Ian Lei}, \bibinfo{person}{Ryan Burnell}, \bibinfo{person}{Libin Bai}, \bibinfo{person}{Anmol Gulati}, \bibinfo{person}{Garrett Tanzer}, \bibinfo{person}{Damien Vincent}, \bibinfo{person}{Zhufeng Pan}, \bibinfo{person}{Shibo Wang}, {et~al\mbox{.}}} \bibinfo{year}{2024}\natexlab{}.
\newblock \showarticletitle{Gemini 1.5: Unlocking multimodal understanding across millions of tokens of context}.
\newblock \bibinfo{journal}{\emph{arXiv preprint arXiv:2403.05530}} (\bibinfo{year}{2024}).
\newblock


\bibitem[Touvron et~al\mbox{.}(2023)]%
        {LLaMA}
\bibfield{author}{\bibinfo{person}{Hugo Touvron}, \bibinfo{person}{Thibaut Lavril}, \bibinfo{person}{Gautier Izacard}, \bibinfo{person}{Xavier Martinet}, \bibinfo{person}{Marie{-}Anne Lachaux}, \bibinfo{person}{Timoth{\'{e}}e Lacroix}, \bibinfo{person}{Baptiste Rozi{\`{e}}re}, \bibinfo{person}{Naman Goyal}, \bibinfo{person}{Eric Hambro}, \bibinfo{person}{Faisal Azhar}, \bibinfo{person}{Aur{\'{e}}lien Rodriguez}, \bibinfo{person}{Armand Joulin}, \bibinfo{person}{Edouard Grave}, {and} \bibinfo{person}{Guillaume Lample}.} \bibinfo{year}{2023}\natexlab{}.
\newblock \showarticletitle{LLaMA: Open and Efficient Foundation Language Models}.
\newblock \bibinfo{journal}{\emph{arXiv preprint arXiv:2302.13971}} (\bibinfo{year}{2023}).
\newblock


\bibitem[Wang et~al\mbox{.}(2025)]%
        {wang2024audiobench}
\bibfield{author}{\bibinfo{person}{Bin Wang}, \bibinfo{person}{Xunlong Zou}, \bibinfo{person}{Geyu Lin}, \bibinfo{person}{Shuo Sun}, \bibinfo{person}{Zhuohan Liu}, \bibinfo{person}{Wenyu Zhang}, \bibinfo{person}{Zhengyuan Liu}, \bibinfo{person}{AiTi Aw}, {and} \bibinfo{person}{Nancy~F Chen}.} \bibinfo{year}{2025}\natexlab{}.
\newblock \showarticletitle{AudioBench: A Universal Benchmark for Audio Large Language Models}.
\newblock \bibinfo{journal}{\emph{NAACL}} (\bibinfo{year}{2025}).
\newblock


\bibitem[Wang et~al\mbox{.}(2021)]%
        {21covost2}
\bibfield{author}{\bibinfo{person}{Changhan Wang}, \bibinfo{person}{Anne Wu}, \bibinfo{person}{Jiatao Gu}, {and} \bibinfo{person}{Juan Pino}.} \bibinfo{year}{2021}\natexlab{}.
\newblock \showarticletitle{CoVoST 2 and Massively Multilingual Speech Translation}. In \bibinfo{booktitle}{\emph{Interspeech}}. \bibinfo{publisher}{{ISCA}}, \bibinfo{pages}{2247--2251}.
\newblock


\bibitem[Wei et~al\mbox{.}(2022)]%
        {22CoT}
\bibfield{author}{\bibinfo{person}{Jason Wei}, \bibinfo{person}{Xuezhi Wang}, \bibinfo{person}{Dale Schuurmans}, \bibinfo{person}{Maarten Bosma}, \bibinfo{person}{Brian Ichter}, \bibinfo{person}{Fei Xia}, \bibinfo{person}{Ed~H. Chi}, \bibinfo{person}{Quoc~V. Le}, {and} \bibinfo{person}{Denny Zhou}.} \bibinfo{year}{2022}\natexlab{}.
\newblock \showarticletitle{Chain-of-Thought Prompting Elicits Reasoning in Large Language Models}. In \bibinfo{booktitle}{\emph{Advances in Neural Information Processing Systems 35: Annual Conference on Neural Information Processing Systems 2022, NeurIPS 2022, New Orleans, LA, USA, November 28 - December 9, 2022}}.
\newblock


\bibitem[Wendler et~al\mbox{.}(2024)]%
        {wendler2024llamas}
\bibfield{author}{\bibinfo{person}{Chris Wendler}, \bibinfo{person}{Veniamin Veselovsky}, \bibinfo{person}{Giovanni Monea}, {and} \bibinfo{person}{Robert West}.} \bibinfo{year}{2024}\natexlab{}.
\newblock \showarticletitle{Do llamas work in english? on the latent language of multilingual transformers}. In \bibinfo{booktitle}{\emph{Proceedings of the 62nd Annual Meeting of the Association for Computational Linguistics (Volume 1: Long Papers)}}. \bibinfo{pages}{15366--15394}.
\newblock


\bibitem[Xue et~al\mbox{.}(2024)]%
        {24echat}
\bibfield{author}{\bibinfo{person}{Hongfei Xue}, \bibinfo{person}{Yuhao Liang}, \bibinfo{person}{Bingshen Mu}, \bibinfo{person}{Shiliang Zhang}, \bibinfo{person}{Qian Chen}, {and} \bibinfo{person}{Lei Xie}.} \bibinfo{year}{2024}\natexlab{}.
\newblock \showarticletitle{E-chat: Emotion-sensitive Spoken Dialogue System with Large Language Models}. In \bibinfo{booktitle}{\emph{{ISCSLP}}}. \bibinfo{publisher}{{IEEE}}.
\newblock


\bibitem[Xue et~al\mbox{.}(2021)]%
        {21mt5}
\bibfield{author}{\bibinfo{person}{Linting Xue}, \bibinfo{person}{Noah Constant}, \bibinfo{person}{Adam Roberts}, \bibinfo{person}{Mihir Kale}, \bibinfo{person}{Rami Al{-}Rfou}, \bibinfo{person}{Aditya Siddhant}, \bibinfo{person}{Aditya Barua}, {and} \bibinfo{person}{Colin Raffel}.} \bibinfo{year}{2021}\natexlab{}.
\newblock \showarticletitle{mT5: {A} Massively Multilingual Pre-trained Text-to-Text Transformer}. In \bibinfo{booktitle}{\emph{Proceedings of the 2021 Conference of the North American Chapter of the Association for Computational Linguistics: Human Language Technologies, {NAACL-HLT} 2021, Online, June 6-11, 2021}}. \bibinfo{publisher}{Association for Computational Linguistics}, \bibinfo{pages}{483--498}.
\newblock


\bibitem[Yue~Yin(2023)]%
        {fujimoto2016reazonspeech}
\bibfield{author}{\bibinfo{person}{Seiji~Fujimoto Yue~Yin, Daijiro~Mori}.} \bibinfo{year}{2023}\natexlab{}.
\newblock \showarticletitle{Reazonspeech: A free and massive corpus for japanese asr}.
\newblock  (\bibinfo{year}{2023}).
\newblock


\bibitem[Yuen et~al\mbox{.}(2024)]%
        {yuen2024internalizing}
\bibfield{author}{\bibinfo{person}{Robin Shing-Hei Yuen}, \bibinfo{person}{Timothy Tin-Long Tse}, {and} \bibinfo{person}{Jian Zhu}.} \bibinfo{year}{2024}\natexlab{}.
\newblock \showarticletitle{Internalizing ASR with Implicit Chain of Thought for Efficient Speech-to-Speech Conversational LLM}. In \bibinfo{booktitle}{\emph{NeurIPS 2024 Workshop}}.
\newblock


\bibitem[Zhao et~al\mbox{.}(2024)]%
        {zhao2024swiftascalablelightweightinfrastructure}
\bibfield{author}{\bibinfo{person}{Yuze Zhao}, \bibinfo{person}{Jintao Huang}, \bibinfo{person}{Jinghan Hu}, \bibinfo{person}{Xingjun Wang}, \bibinfo{person}{Yunlin Mao}, \bibinfo{person}{Daoze Zhang}, \bibinfo{person}{Zeyinzi Jiang}, \bibinfo{person}{Zhikai Wu}, \bibinfo{person}{Baole Ai}, \bibinfo{person}{Ang Wang}, \bibinfo{person}{Wenmeng Zhou}, {and} \bibinfo{person}{Yingda Chen}.} \bibinfo{year}{2024}\natexlab{}.
\newblock \bibinfo{title}{SWIFT:A Scalable lightWeight Infrastructure for Fine-Tuning}.
\newblock
\showeprint[arxiv]{2408.05517}~[cs.CL]


\bibitem[Zhu et~al\mbox{.}(2024)]%
        {24QuestionTranslation}
\bibfield{author}{\bibinfo{person}{Wenhao Zhu}, \bibinfo{person}{Shujian Huang}, \bibinfo{person}{Fei Yuan}, \bibinfo{person}{Shuaijie She}, \bibinfo{person}{Jiajun Chen}, {and} \bibinfo{person}{Alexandra Birch}.} \bibinfo{year}{2024}\natexlab{}.
\newblock \showarticletitle{Question Translation Training for Better Multilingual Reasoning}. In \bibinfo{booktitle}{\emph{Findings of the Association for Computational Linguistics, {ACL} 2024, Bangkok, Thailand and virtual meeting, August 11-16, 2024}}. \bibinfo{publisher}{Association for Computational Linguistics}, \bibinfo{pages}{8411--8423}.
\newblock


\end{thebibliography}










\end{document}